\documentclass[aps,prl,twocolumn,superscriptaddress,nofootinbib]{revtex4-1}
\usepackage{amsmath}
\usepackage{graphicx}
\usepackage{sistyle}
\usepackage{gensymb}
\usepackage{bm}
\usepackage{natbib}
\usepackage{dsfont}
\usepackage{epsfig}
\usepackage[normalem]{ulem}

\newcommand{\be}{\begin{eqnarray}}
\newcommand{\ee}{\end{eqnarray}}



\usepackage{color}

\begin{document}

~~~~~~~~~~~~~~~~~~~~~~~~~~ {FERMILAB-PUB-20-282-AE-T}

\medskip

\title{Search for composite dark matter with optically levitated sensors}

\author{Fernando Monteiro}
\email{fernando.monteiro@yale.edu}
\author{Gadi Afek}
\affiliation{Wright Laboratory, Department of Physics, Yale University, New Haven, CT, USA}
\author{Daniel Carney}
\affiliation{Joint Center for Quantum Information and Computer Science/Joint Quantum Institute, 
University of Maryland/NIST, College Park/Gaithersburg, MD, USA}
\affiliation{Fermi National Accelerator Laboratory, Batavia, IL, USA}
\author{Gordan Krnjaic}
\affiliation{Fermi National Accelerator Laboratory, Batavia, IL, USA}
\affiliation{Kavli Institute for
Cosmological Physics, Universtiy of Chicago, Chicago, IL, USA}
\author{Jiaxiang Wang} 
\author{David C. Moore}
\affiliation{Wright Laboratory, Department of Physics, Yale University, New Haven, CT, USA}

\begin{abstract}
Results are reported from a search for a class of composite dark matter models with feeble, long-range interactions with normal matter. We search for impulses arising from passing dark matter particles by monitoring the mechanical motion of an optically levitated nanogram mass over the course of several days.  Assuming such particles constitute the dominant component of dark matter, this search places upper limits on their interaction with neutrons of $\alpha_n \leq 1.2 \times 10^{-7}$ at 95\% confidence for dark matter masses between 1--10~TeV and mediator masses $m_\phi \leq 0.1$~eV.  Due to the large enhancement of the cross-section for dark matter to coherently scatter from a nanogram mass ($\sim 10^{29}$ times that for a single neutron) and the ability to detect momentum transfers as small as $\sim$200 MeV/c, these results provide sensitivity to certain classes of composite dark matter models that substantially exceeds existing searches, including those employing kg-scale or ton-scale targets. Extensions of these techniques can enable directionally-sensitive searches for a broad class of previously inaccessible heavy dark matter candidates. 
\end{abstract}

\maketitle


There is compelling evidence for the existence of dark matter (DM) from multiple independent cosmological and astrophysical sources~\cite{Primack:2015kpa, Aghanim:2018eyx, Massey:2010hh, randall:2007ph, Sofue:2000jx, Jedamzik:2009uy}. However, its detection in terrestrial experiments has eluded searches to date, and remains among the highest priorities in fundamental physics. The most sensitive laboratory detection strategies typically involve searching for subatomic particle recoils if the DM mass, $M_X$, is sufficiently large $(M_X \gtrsim \rm eV)$
 \cite{Schumann:2019eaa} or DM-photon~\cite{Graham:2015ouw} or DM-phonon~\cite{Knapen:2016cue,Knapen:2017ekk,Cox:2019cod} conversion for smaller masses $(M_{X} \ll \rm eV)$. Both strategies aim to observe tiny energy transfers between DM and microscopic internal degrees of freedom of a massive (or large volume) detector.
 
Recent work has suggested the possibility that macroscopic force sensors can be used to probe long-range interactions between dark and visible matter, including---in principle---those due to gravity alone~\cite{carney:2019_1,PhysRevD.98.083019,PhysRevD.99.023005}.  While realizing such ambitious experiments would require substantial advances beyond the current state-of-the-art, similar concepts to search for DM that may interact via stronger long-range interactions are already feasible.  

In this {\it Letter}, we search for passing DM particles by monitoring impulses delivered to a macroscopic sensor through its center-of-mass (COM) motion. 
The ability to detect tiny momentum transfers to nanogram-scale masses is enabled by the extreme sensitivity of recently developed levitated optomechanical systems. 
Techniques to trap micron or sub-micron sized masses via optical~\cite{Ashkin:1971,Li:2011,Gieseler:2012}, magnetic~\cite{DUrso:2018,Ulbricht:2019,BrianD'Urso_2020,Gieseler:2020}, or radio-frequency~\cite{Dania:2020kzl,2020JPhD...53q5302B,Millen:2018,Barker:2015} fields have progressed substantially in the last decade~\cite{Millen:2020review}.  Past work has demonstrated the ability to cool particles with $\sim$fg masses to $\mu$K effective temperatures~\cite{Novotny2019, Aspelmeyer:2019, Novotny:2020}. Recent extensions of such cooling to masses as large as 1~ng~\cite{acceleration2020} is key to enabling the DM searches presented here. While such techniques may also enable tests of quantum mechanics using massive objects~\cite{Vinante2019, Novotny2019, Aspelmeyer:2019, Novotny:2020}, precise micron-scale accelerometers and force sensors~\cite{gieseler:2013,Ranjit:2016,Hempston:2017,Novotny_static:2018,Ranjit:2015,Rider:2017,Blakemore_3D_microscope:2019,Acceleration_2017, acceleration2020}, and searches for new fundamental interactions~\cite{Moore:2014, Rider:2016}, here we provide an initial demonstration of their ability to detect small recoils, including those that might arise from DM particles interacting via a long-range force.

Motivated by recent theoretical developments \cite{Krnjaic:2014xza,Gresham:2017zqi,hardy2014big,Hardy:2015boa,Ibe:2018juk,Grabowska:2018lnd}, we consider models of composite DM ``nuggets" that interact with visible matter through a classical Yukawa potential mediated by a light force carrier, $\phi$, with mass $m_\phi \lesssim$ eV:
\be
\label{yukawa}
V(r) =   \frac{ \alpha}{r}e^{-r/ \lambda} ~~,~~ 
 \alpha \equiv \frac{(N_d g_d) (N_n g_n)}{4\pi} ,
\ee
where 
$g_d$ is the coupling of $\phi$ to DM nugget constituents, 
$g_n$ is its coupling to neutrons, 
$N_d \gg 1$ is the number of constituents in the nugget, and $N_n \sim 3\times10^{14}$ is the number of neutrons in the sensor. The range of the force is $\lambda \equiv m^{-1}_\phi \simeq 2 \, {\rm \mu m} \times (0.1\ \mathrm{eV}/m_\phi)$, and we denote the coupling of the entire DM nugget to a single neutron as $\alpha_n = \alpha/N_n$. 
Here, and when specifying particle masses or momentum transfers in this paper, natural units are used with $\hbar = c = 1$. Other experimental parameters are reported in SI units for clarity. While the neutron coupling to light mediators, $g_n$, is strongly constrained by fifth force searches and equivalence principle tests~\cite{Murata_2015,EotWash2020}, $g_d$ is considerably less constrained. Couplings for which $g_d \gg g_n$ are typically required to produce observable signals. 

The sensor consists of an SiO$_2$ sphere with diameter $d_\text{sph} = 10.3\pm1.4$~$\mu$m and mass density $\rho_\text{sph} = 1.8$~g/cm$^3$~\cite{Acceleration_2017}, which is optically levitated in high vacuum. A detailed description of the trapping setup is given in~\cite{acceleration2020}. Active feedback is used to cool the sphere's COM motion in all translational degrees of freedom to an effective temperature, $T_{eff} \approx 200\ \mu$K, which simulations indicate provides the optimal impulse sensitivity for the measured force noise, $\sqrt{S_{FF}}\approx 1$~aN$/\sqrt{\text{Hz}}$~\cite{acceleration2020}. Data were acquired during a 7~day period between June 15--21, 2020. Prior to beginning data acquisition, the sphere was optically trapped at $\lesssim 5\times10^{-7}$~mbar 
and its net electric charge neutralized~\cite{Moore:2014,acceleration2020}, remaining zero (with no spontaneous charging) throughout data acquisition.  

\begin{figure}[t]
    \centering
    \includegraphics[width=\columnwidth]{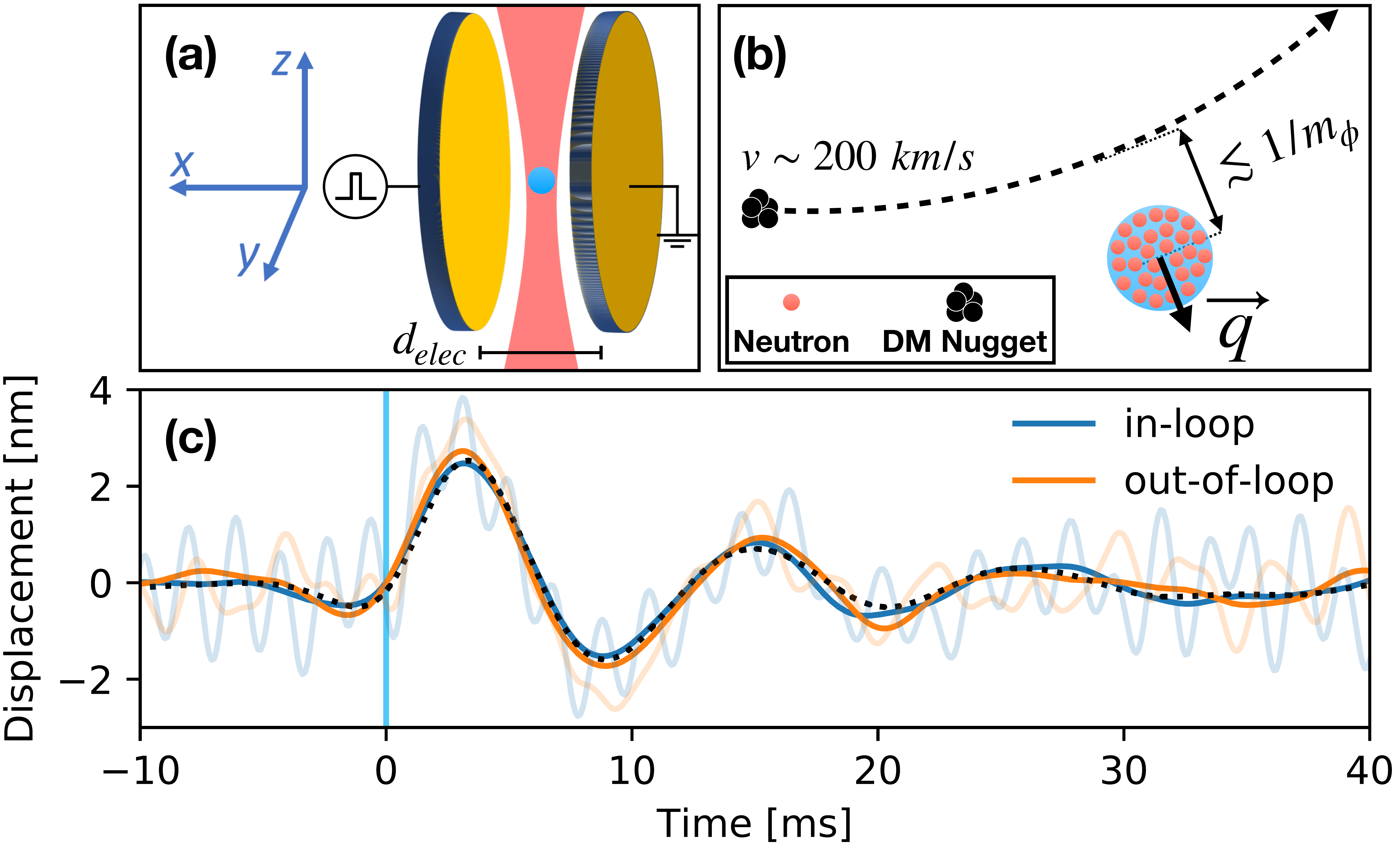}
    \caption{ Schematics of: (a) the levitated sphere and calibration electrodes, and (b) a DM nugget coherently scattering from a sphere via a light mediator, producing a momentum transfer $\vec{q}$. (c) Example 4.8~GeV impulse produced by applying a pulsed electric field at $t=0$ to a sphere with charge $-1\ e$. The raw waveform with minimal filtering (light, solid), filtered waveform (dark, solid), and filtered template (dashed) are shown.}
    \label{fig:impulse}
\end{figure}

This work considers only motion of the sphere in the $x$-direction [see Fig.~\ref{fig:impulse}(a)], since the impulse response could be directly calibrated using existing electrodes surrounding the trap.  Upgrades to add additional electrodes can allow accurate calibration of the sphere's 3D motion~\cite{Blakemore_3D_microscope:2019,Kawasaki:2020oyl}.
The sphere's $x$-position was measured using two independent sensors: one within the feedback loop (``in-loop'') and one utilizing a separate imaging beam and photodiode (``out-of-loop'')~\cite{acceleration2020}. Data from a commercial accelerometer (Wilcoxon 731A/P31) positioned just outside the vacuum chamber were also recorded. Data from all sensors were continuously acquired in $\sim 105$~s long data files ($2^{20}$ samples at a sampling rate of 10~kHz). Additional data were taken
to calibrate impulse amplitudes and measure selection and reconstruction efficiencies during dedicated runs performed at the beginning, middle, and end of the acquisition period. 

To search for candidate impulse events, waveforms in each data file are first filtered to remove narrow lines and out-of-band noise, while preserving the majority of the signal around the resonance frequency $f_0\sim85$~Hz.
The same filter is then applied to a signal template constructed from the expected impulse response of a damped harmonic oscillator, using $f_0$ and the damping coefficient, $\Gamma_0\sim35$~Hz.  These parameters were determined from the calibration data and stable within 5\% and 10\%, respectively, throughout the acquisition period. After filtering, the template and waveform are cross-correlated, and local extrema (i.e., candidate impulses) are identified in the correlated data, for which the amplitude, time, and $\chi^2$ goodness-of-fit statistic are recorded.  This reconstruction is performed for the in-loop and out-of-loop waveforms in calibration and DM-search data.

Passing DM particles [Fig.~\ref{fig:impulse}(b)] will impart an impulse over a time $\Delta t \sim b_\text{max}/v$, where $b_\text{max}$ is the largest impact parameter at which a sufficiently large signal is produced and $v \sim 200$~km/s is the DM velocity.  At all DM masses and couplings considered here, $b_\text{max} \lesssim 1$~mm, and the resulting impulses are essentially instantaneous ($\Delta t \lesssim 5\ \mathrm{ns}$) relative to the $\sim$ms sphere response time. To mimic this signal in the calibration data, a net electric charge of $-1\ e$ was added to the sphere and a sequence of square voltage pulses (of length $\Delta t = 100$~$\mu$s) with fixed amplitudes ranging between 20~V and 1.28~kV was applied to the calibration electrodes, which had measured spacing $d_\text{elec} = 3.99 \pm 0.05$~mm. The impulse time is sufficiently long to avoid distortion by the high voltage amplifier (Trek 2220), but remains short compared to the sphere response time. 

Each calibration run consisted of $\sim 200$ impulses for each of 7 amplitudes in the range 0.15--9.6~GeV. The applied impulses span the analysis range considered here and provide a direct calibration of the reconstructed impulse amplitudes in the DM-search data, with relative amplitude uncertainty of 1.3\% dominated by the uncertainty on $d_\text{elec}$. This calibration technique avoids uncertainties related to the sphere mass and accounts for small time variations in $f_0$ and $\Gamma_0$.  Figure~\ref{fig:impulse}(c) shows an example of the calibrated response.  Prior to calibration, the reconstructed amplitudes were linear within 1\% over the range from 1--10~GeV.  At amplitudes $\lesssim 1$~GeV the calibration removes non-linearity due to template search bias~\cite{Moore:2012gya}.  

Data selection cuts were applied to avoid spurious signals from environmental noise.  First, a significant increase in the number of noise-like events was observed when someone was present in the lab. A ``lab entry'' cut was applied to exclude such periods based on a detailed lab-access log, which removed 0.82~days (14\%) of livetime. During these noisy periods, the vibrational impulses were found to be both correlated in time (i.e., a short sequence of large impulses would typically be recorded, rather than single, isolated events), and to correlate with those measured by the commercial accelerometer.  These observations motivated two additional event selection cuts.  An ``accelerometer cut'' was applied to exclude data files for which the maximum deviation in the filtered accelerometer time stream was $>2.5 \sigma$ larger than the mean of the distribution, removing 2.6\% of livetime remaining after the lab entry cut.  In addition, an ``anti-coincidence cut'' was used to exclude any 1~s long time period where 2 or more events were reconstructed with amplitudes larger than 1~GeV, further reducing livetime by 0.2\% after the previous cuts. The estimated signal efficiency of the anti-coincidence cut due to random coincidence of the observed rate of isolated impulses in the dataset with a DM signal assumed to be uniformly distributed in time is $>99.5$\%.  After all selection cuts, the remaining livetime for the DM search is 4.97~days.

\begin{figure}[t]
    \centering
    \includegraphics[width=\columnwidth]{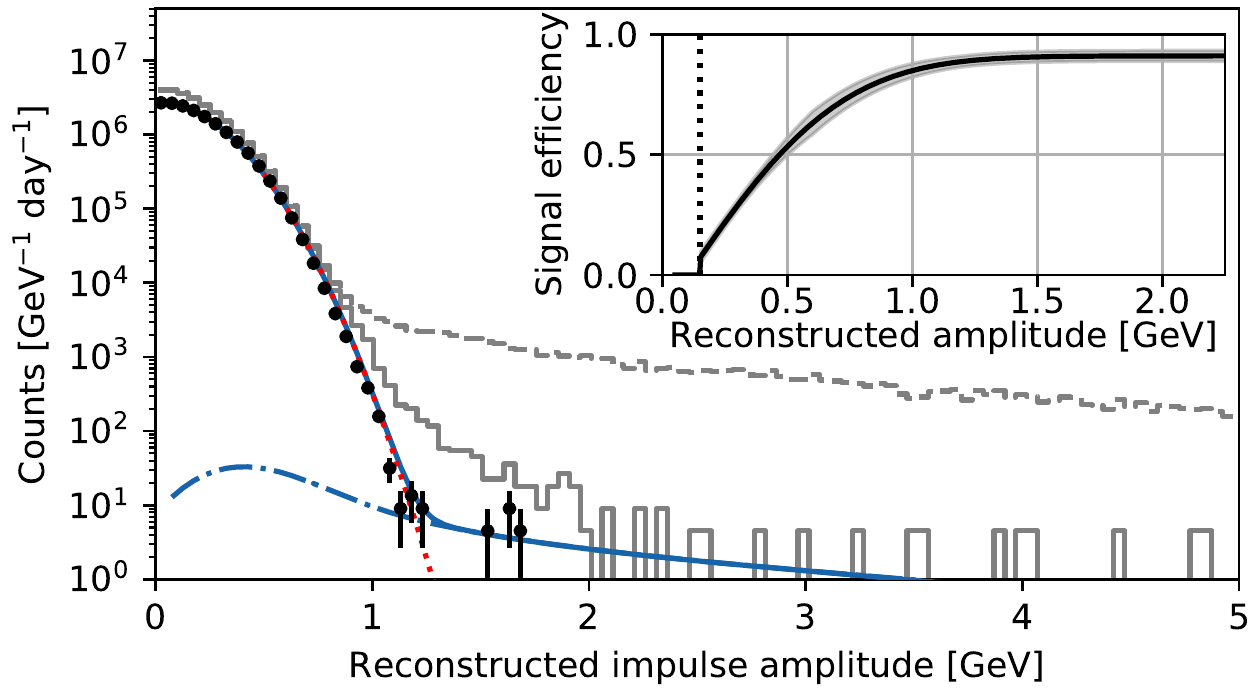}
    \caption{Measured rate of reconstructed impulses after all cuts (black points), compared to the spectrum with only livetime selections applied (gray, solid) and with no cuts applied (gray, dashed).  The Gaussian background (red, dotted), DM signal (blue, dot-dashed), and sum of background and signal (blue, solid) are also shown at the 95\% CL upper limit, $\alpha_n={8.5\times10^{-8}}$, for $M_X = 5\times 10^3$~GeV, $m_\phi = 0.1$ eV, and $f_X = 1$.  (Inset) Overall signal efficiency versus amplitude (black) and estimated error (gray band) above the analysis threshold, $q_\text{thr}=0.15$~GeV (dotted). 
    }
    \label{fig:spectrum}
\end{figure}

For events passing the livetime selection, two event-level quality cuts were applied.  First, the in-loop and out-of-loop amplitudes were required to be consistent within the combined resolution of both sensors. The signal efficiency for this cut was measured from the calibration pulses to be $95.0 \pm 1.2$\%, independent of amplitude.  Second, the $\chi^2$ statistic between the waveform and best-fit template was required to be consistent with the distribution for calibration pulses. The cut threshold was empirically set to accept an equal fraction of calibration events at each amplitude, resulting in a measured efficiency of $95.9\pm1.8$\%.  
Finally, the calibration was used to determine the impulse detection efficiency versus amplitude, which provides the dominant inefficiency for reconstructing small impulses.  The detection efficiency was measured from calibration data by counting the fraction of applied impulses at each amplitude for which a reconstructed impulse was detected, after correcting for the rate of accidental coincidences from noise.  The detection efficiency is measured to be $8.0 \pm 1.1$\% at the analysis threshold of 0.15~GeV and rises to $\sim$100\% for impulses larger than 0.9~GeV.  The overall signal efficiency estimated from the combination of the detection and cut efficiencies is shown in the inset of Fig.~\ref{fig:spectrum}.

The distribution of reconstructed impulses is shown in Fig.~\ref{fig:spectrum}, both before and after applying livetime and quality cuts.  For impulses below 1.2~GeV, the data are consistent with a Gaussian distribution resulting from the random reconstruction of noise events near threshold.  After all cuts, a non-Gaussian tail of 4 events is observed between 1.2--1.7~GeV, with no additional events from 1.7--10~GeV. While the simple analysis performed here cannot distinguish such events from a DM signal, their distribution is similar to the much higher rate of background-like events removed by the livetime selection and quality cuts.  Given this similarity, we do not report a best-fit DM signal and instead set limits on the coupling of DM particles to neutrons in the sphere under the assumption that such events could arise either from DM-induced signals or backgrounds.  We note that the optomechanical sensors used here are directionally sensitive, and future analyses searching for diurnal modulation in the distribution of recoil directions could definitively separate a DM-induced signal from backgrounds~\cite{PhysRevD.37.1353,Ahlen:2009ev}.

To determine the upper limit on the DM-neutron coupling, $\alpha_n$, for a given mediator mass, $m_\phi$, and dark matter mass, $M_X$, a profile-likelihood based hypothesis test is used~\cite{Rolke:2004mj}.  The binned negative log-likelihood (NLL) is calculated for the data and a model consisting of a Gaussian background plus the calculated differential rate of DM-induced impulses:
\begin{equation}
\frac{dR}{dq}(\alpha_n; M_X, m_{\phi}) = \frac{f_X \rho_X}{M_X} \int\limits{dv\, v f(v) \frac{d\sigma}{dq}}    
\end{equation}
where $\rho_X = 0.3$~GeV/cm$^3$ is the local DM density~\cite{Tanabashi:2018oca}, for which the composite DM candidate of interest accounts for a fraction $f_X$ of the total density, $v$ is the DM velocity with distribution, $f(v)$, and the differential cross-section, $\frac{d\sigma}{dq}$,
is determined numerically for classical scattering from the potential in Eq.~\ref{yukawa}~\cite{goldstein2002classical}, generalized to a uniform density sphere of diameter $d_\text{sph}$ and projected onto the $x$-direction. The ``standard halo model'' (SHM) for $f(v)$ is assumed, with $v_0 = 220$~km/s~\cite{1986MNRAS.221.1023K}, escape velocity $v_\text{esc} = 544$~km/s~\cite{10.1111/j.1365-2966.2007.11964.x}, and average Earth velocity $v_{e} = 245$~km/s~\cite{10.1111/j.1365-2966.2010.16253.x}. The minimum velocity to produce a recoil above threshold is $v_\text{min} = q_\text{thr}/(2 M_X)$ for the analysis threshold, $q_\text{thr} = 0.15$~GeV.  The upper limit on the analysis range is 10~GeV.  The differential rate is corrected by the signal efficiency [Fig.~\ref{fig:spectrum} (inset)] and convolved with a Gaussian of width $\sigma = 0.17$~GeV to account for the momentum resolution.
Nuisance parameters account for systematic errors and backgrounds, including: the amplitude of the Gaussian background; a multiplicative scaling of the momentum; and a multiplicative scaling of $N_n$. While the background amplitude is allowed to float freely, the latter two parameters are constrained by Gaussian terms in the NLL with unity means and $\sigma = 1.3$\% and 35\%, corresponding to the uncertainties for $d_\text{elec}$ and $N_n \propto d_\text{sph}^3$, respectively. 

\begin{figure}[t]
    \centering
    \includegraphics[width=\columnwidth]{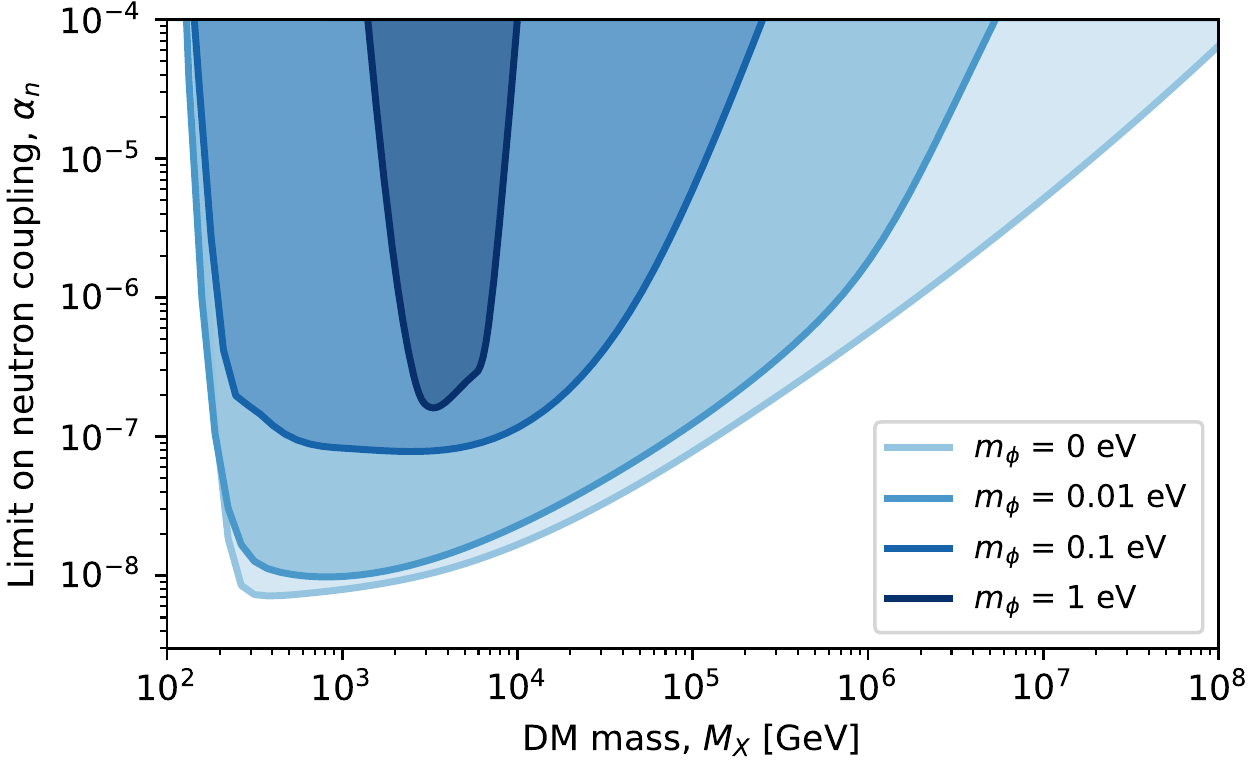}
    \caption{95\% CL upper limits on the DM-neutron coupling, $\alpha_n$, versus DM mass, $M_X$, for several example values of mediator mass, $m_\phi$, assuming $f_X = 1$.}
    \label{fig:limits_model_indep}
\end{figure}

The resulting 95\% CL upper limits on $\alpha_n$ are shown in Fig.~\ref{fig:limits_model_indep}. For $m_\phi \ll 1/b_\text{max}$, the limits converge to those for a massless mediator.  For $1/b_\text{max} \lesssim m_\phi \lesssim 1/d_\text{sph}$, sensitivity to $\alpha_n$ is reduced due to the reduction in cross section to $\sim m_\phi^{-2}$, and further reduced for $m_\phi \gtrsim 1/d_\text{sph}$ by the form-factor suppression from interaction of the DM with only a fraction of the neutrons in the sphere.  In all cases, the limits become weaker at large $M_X$ due to the reduced DM number density and at small $M_X$ due to the momentum threshold.

While the results in Fig.~\ref{fig:limits_model_indep} apply for any DM model interacting with neutrons via the generic potential in Eq.~\ref{yukawa}, they can also be translated to a specific microscopic model. As an example, we consider bound states of asymmetric DM~\cite{PhysRevD.100.035025,Grabowska:2018lnd} in which composite DM nuggets of total mass $M_X$ can be formed from a large number ($N_d > 10^4$) of lighter constituents, each with mass $m_d$.  Recent studies indicate that such composite particles provide viable DM candidates and could be formed in the early universe at the required densities to constitute some, or all, of the relic DM density~\cite{Krnjaic:2014xza,Gresham:2017zqi,hardy2014big,Hardy:2015boa,Ibe:2018juk,Grabowska:2018lnd}. 

\begin{figure}[t]
    \centering
    \includegraphics[width=\columnwidth]{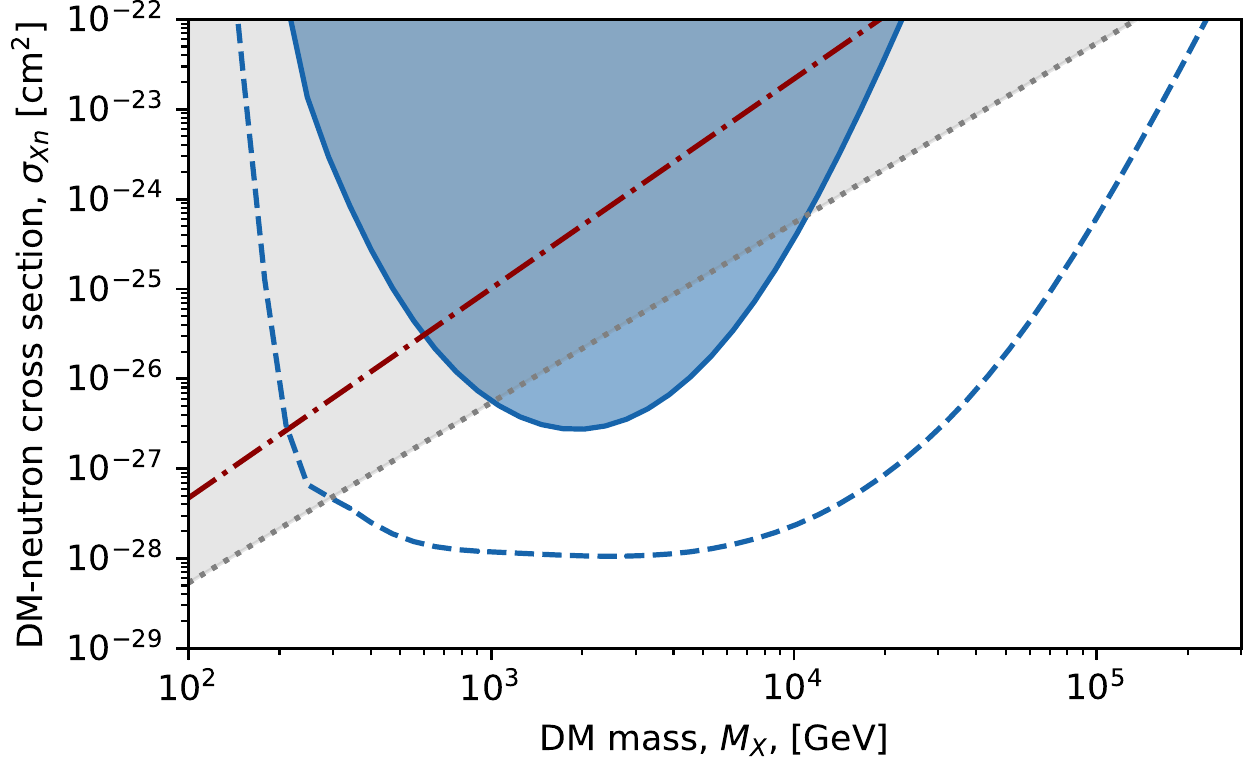}
    \caption{
    Upper limits on the equivalent DM-neutron scattering cross-section for a point-like nugget, $\sigma_{Xn} \equiv 4\pi \alpha_n^2 \mu_{Xn}^2/q_0^4$~\cite{PhysRevD.100.035025}, versus $M_X$, for the model described in the text with $f_X = 0.1$ (solid) and $f_X = 1$ (dashed). Here $\sigma_{Xn}$ is evaluated for $m_d = 1$~keV, $m_\phi = 0.1$~eV, and at a reference momentum of $q_0 = m_n v_0$ where $m_n$ is the neutron mass and $\mu_{Xn}$ is the DM-neutron reduced mass. Model-dependent fifth-force constraints~\cite{Murata_2015,EotWash2020} (dotted) are also shown, assuming $g_d \approx 1$. Due to sharp DM nugget form-factor suppression in the parameter space chosen here, existing detectors searching for $\sim$eV--keV scale NRs~\cite{PhysRevD.97.022002,PhysRevD.100.102002,PhysRevD.94.082006,PhysRevLett.121.061803,PhysRevLett.121.111302, PhysRevLett.118.021303,PhysRevLett.119.181302,PhysRevLett.121.081307} only  constrain $\sigma_{Xn} \gg 10^{-22}$~cm$^2$. The results reported here exceed even the projected sensitivity of a $\sim$kg-yr exposure of an ambitious future detector with NR threshold as low as 1~meV (dot-dashed, see, e.g., \cite{Fink:2020noh,PhysRevD.100.092007,PhysRevD.100.035025,PhysRevLett.123.151802}). CMB limits on DM-baryon interactions assume a coupling to protons, which is 
    model-dependent and need not apply here \cite{Boddy:2018wzy}, although the $f_X = 1$ region is expected to be excluded by DM self-interaction bounds~\cite{10.1093/mnras/stt2097,PhysRevD.100.035025}, which do not apply for $f_X \lesssim 0.1$.
    }
    \label{fig:limits_composite}
\end{figure}

Example constraints from this search for $m_\phi = 0.1$~eV, $m_d = 1$~keV, and $f_X = (0.1, 1)$ are shown in Fig.~\ref{fig:limits_composite}. In contrast to nuclear recoils (NR) from nuggets with these parameters~\cite{PhysRevD.100.035025}, screening of the interaction within the nugget has negligible effect on $d\sigma/dq$ regardless of $g_d$ since the geometric cross section of the nugget is much smaller than the total cross section, for all $M_X$ considered.  For these parameters, bounds on the DM-DM scattering cross-section~\cite{10.1093/mnras/stt2097} are expected to prevent such nuggets from providing the dominant component of DM, but cannot constrain such models if they provide only a subcomponent of the total relic density, with $f_X \lesssim 0.1$~\cite{PhysRevD.100.035025}.  In such models, which typically contain a complex dark sector and a correspondingly complex formation history, production of a subcomponent of such composite particles is generically possible, similar to the wide range of composite particles formed in the visible sector. Use of the SHM allows direct comparison of the results presented here to the projected sensitivity of existing and future detectors in previous work~\cite{PhysRevD.100.035025}. However, if deviations from the SHM arise, e.g., from DM self-interactions, then the derived limits could be modified. For example, the limits would generally be strengthened if $f_X \rho_X$ were larger than assumed in the SHM, or if the local velocity distribution were shifted towards lower velocities.

These results---using only a single, nanogram-mass sphere and less than a week of livetime---already provide many orders of magnitude more sensitivity to DM interactions in these models than existing direct detection searches. Large detectors searching for DM-induced NRs using cryogenic calorimeters~\cite{PhysRevD.97.022002,PhysRevD.100.102002}, semiconductors~\cite{PhysRevD.94.082006,PhysRevLett.121.061803}, or liquid noble targets~\cite{PhysRevLett.121.111302, PhysRevLett.118.021303,PhysRevLett.119.181302,PhysRevLett.121.081307} do not significantly constrain these models due to the low probability of producing events above their $\sim$eV to keV scale energy thresholds.  In contrast, the techniques presented here (similar to other proposed techniques utilizing collective excitations of many atoms, e.g.,~\cite{PhysRevD.100.035025,Knapen:2016cue,Griffin:2018bjn}) take advantage of the large enhancement in cross-section from scattering coherently from a nanogram mass and ability to detect momentum transfers as small as $\sim 0.2$~GeV, corresponding to a recoil energy of the sphere's COM motion of $\sim$30~neV. For sufficiently massive mediators and light constituents, such as the parameters shown in Fig.~\ref{fig:limits_composite}, and assuming $g_d \approx 1$, these results extend between 1--3 orders of magnitude beyond stringent constraints from fifth-force bounds on $g_n$, if such particles make up a fraction between $f_X = 0.1$--1 of the relic DM density.  

In summary, this work searches for previously unexplored classes of DM particles interacting with normal matter through a long-range force using nanogram-scale, optomechanical sensors.  With only a few days of livetime, the techniques presented here extend sensitivity by many orders-of-magnitude beyond traditional WIMP detectors for the benchmark model considered, and surpass stringent, but model-dependent constraints on DM-neutron interactions arising from fifth-force experiments.  These results provide an initial experimental demonstration of a general class of new techniques to search for DM using optomechanical sensors~\cite{carney:2019_1,PhysRevD.98.083019,PhysRevD.99.023005,Carney_white_paper,Cheng:2019vwy, Ulbricht2015}, and future searches with optimized systems are expected to substantially exceed the sensitivities obtained here by reaching lower thresholds and longer livetimes~\cite{Carney_white_paper}.  Large arrays of sensors could allow track-like signals from DM particles to be reconstructed~\cite{carney:2019_1}. Finally, if DM-induced collisions were detected with optomechanical sensors, ``smoking-gun'' evidence for the origin of such signals could be confirmed through their natural directional sensitivity~\cite{PhysRevD.37.1353,Ahlen:2009ev}.  

We would like to thank the Gratta group at Stanford for useful discussions related to the apparatus used in this work. This work is supported, in part, by the Heising-Simons Foundation, NSF Grant PHY-1653232, and the Alfred P. Sloan Foundation. Initial discussions leading to this work occurred at the Aspen Center for Physics, which is supported by NSF Grant PHY-1607611. This manuscript has been authored by Fermi Research Alliance, LLC under Contract No. DE-AC02-07CH11359 with the U.S. Department of Energy, Office of High Energy Physics. 

\bibliographystyle{apsrev4-1}
\bibliography{references}{}

\end{document}